\documentclass[letter]{aa}
\usepackage{graphicx}
\usepackage{txfonts}

\begin{document}

\title{The dark matter halo of NGC\,1399 - CDM or MOND?
\thanks{Based on observations made with ESO Telescopes at the  Paranal Observatories under program ID 70.B-0174.}\fnmsep 
\thanks{Based on observations obtained at the Gemini Observatory, which
is operated by the Association of Universities for Research in
Astronomy, Inc., under a cooperative agreement with the NSF on behalf
of the Gemini partnership: the National Science Foundation (United
States), the Particle Physics and Astronomy Research Council (United
Kingdom), the National Research Council (Canada), CONICYT (Chile), the
5Australian Research Council (Australia), CNPq (Brazil) and CONICET
(Argentina).}}
\subtitle{}
\author{T. Richtler
           \inst{1}
           \and
       Y. Schuberth
           \inst{1,2}
            \and
        M. Hilker
         \inst{3}
         \and
         B. Dirsch
         \inst{1}
         \and
         L. Bassino
         \inst{4}
         \and 
         A.J. Romanowsky
       \inst{1,5}
}

\offprints{T. Richtler}
\institute{Departamento de F\'{\i}sica, Casilla 160-C, Universidad de Concepci\'on, Concepci\'on, Chile (contact adress for BD)
        \and
         Argelander Institut f\"ur Astronomie, Auf dem H\"ugel 71, 53121 Bonn, Germany
         \and
          European Southern Observatory, Karl-Schwarschildstr.2, Garching, Germany
         \and 
Facultad de Ciencias Astron\'omicas y Geof\'{\i}sicas,
       Universidad Nacional de La Plata,
       Paseo del Bosque S/N, 1900-La Plata,\\
       Argentina; and IALP-CONICET
       \and
       UCO/Lick Observatory, University of Santa Cruz, California, 95064, USA
}
\date{}
\abstract{Central galaxies in galaxy clusters may be key discriminants in the 
competition between the cold dark matter (CDM) paradigm and modified Newtonian dynamics (MOND). }
{We investigate the dark  halo of NGC\,1399, the central galaxy of the Fornax cluster, out to a galactocentric distance of 80\,kpc.}
{The data base consists of  656 radial velocities of globular clusters obtained with MXU/VLT and
GMOS/Gemini, which is the largest sample so far for any  galaxy. We performed a Jeans analysis for a non-rotating isotropic model.}{An NFW halo with the parameters $r_s = 50\,\rm{kpc}$ and $\varrho_s = 0.0065 M_\odot/\rm{pc}^3$ provides a
good description of our data, fitting well to the X-ray mass.  More massive halos are also permitted that agree with the mass of the Fornax cluster
as derived from galaxy velocities. 
 We compare this halo with the expected MOND models under isotropy
and find that additional dark matter on the order of the stellar mass is needed to get agreement.
A fully radial infinite globular cluster system would be needed to change this conclusion. 
} 
{Regarding CDM, 
we cannot draw firm conclusions. To really constrain a cluster wide halo, more data covering
a larger radius are necessary. The MOND result appears as a small-scale
 variant
of the finding that MOND in galaxy clusters still needs dark matter.}
\keywords{galaxies: elliptical and lenticular, cD ---
galaxies: kinematics and dynamics --- galaxies:individual:NGC\,1399}

\maketitle

\section{Introduction}

The dark matter halos of early-type galaxies are  investigated less well
than those
of spiral galaxies (for a recent review, see Romanowsky \cite{romanowsky06}).
 Globular clusters (GCs) and planetary nebulae are  the
main stellar dynamical tracers, while X-rays can be used at even larger galactocentric distances for a gas dynamical approach.
The reliability of a deduced mass profile depends strongly on the number of dynamical probes
available. Regarding GCs, it is therefore natural that the rich GC systems of bright elliptical galaxies have so far been  the preferred targets for such studies,
for example M87 (C{\^o}t{\'e} et al.~\cite{cote01}), NGC\,4472 (C{\^o}t{\'e} et al.~\cite{cote03}), and NGC\,4636 (Schuberth et al.~\cite{schuberth06}). The largest sample until
 today has been measured for NGC\,1399, the central galaxy of the Fornax galaxy
 cluster. Richtler et al.~(\cite{richtler04}, Paper\,I) analyzed the velocities
 of about 450 GCs out to a galactocentric distance of 40\,kpc and found an approximately 
 constant velocity dispersion. The corresponding dark matter halo was
found to be consistent with an NFW-halo (see section 5 for the numerical values). 
The relatively poor knowledge that we have until now regarding the dark halos of early-type galaxies,
does not indicate a common behavior such as  the constant rotation curves found in spiral galaxies. While the inferred circular velocities of NGC\,1399 or NGC\,4636 appear to be flat, they
seem to rise for M87, and most strikingly so for NGC\,6166 (Kelson et al.~\cite{kelson02}), the central galaxy in Abell 2199. In elliptical galaxies of lower luminosity, the projected velocity dispersion
can even be falling (Romanowsky et al.~\cite{romanowsky03}), perhaps including the circular velocity.
 One may speculate that
these relate to the environment. The dark halos of central galaxies might be determined by the dark halos of their host galaxy clusters rather than being halos belonging to  individual galaxies. However, a larger sample of well-studied galaxies is necessary to 
arrive at firm conclusions.\\
We have now augmented the number of GC velocities around NGC\,1399 to nearly 660 reaching a galactocentric distance of 80\,kpc, which is an unprecedently large and extended sample.
In this \emph{Letter}, we present the implications of this sample for the dark halo,
updating the conclusions of Paper\,I.
 Moreover, we discuss the dark halo in the context of Modified Newtonian Dynamics (MOND) (e.g. Milgrom \cite{milgrom83}, Sanders \& McGaugh \cite{sanders02}). The detailed description of the observations, the reduction, and the analysis will be presented in a
forthcoming paper (Schuberth et al., in prep), taking into account the population substructure of the GC system as well as more anisotropic models.\\  
In the following, we adopt a distance of 19\,Mpc for NGC\,1399. 

\section{Observations}
Our new data were 
obtained with the multi--object--spectrographs FORS2/MXU and GMOS at
the VLT and Gemini--South observatories, respectively. 
The GC candidates were selected from the wide--field
photometry by Dirsch et al.~(\cite{dirsch03}). 
The spectroscopic observations of ten MXU masks
in seven fields were carried out in visitor mode during three nights
(1--3 December 2002) at the European Southern Observatory Very Large
Telescope (VLT) facility on Cerro Paranal, Chile.
The observations  with the Gemini Multi--Object Spectrograph (GMOS) on Gemini--South  were carried out in queue mode in November 2003
and December 2004. A total of ten spectroscopic masks in five fields
were observed.
We then merged the velocity catalogues with the previous velocities of Paper\,I after having
revised the old spectra by cross correlation methods with the same templates used for the new data. 
The final velocity database comprises 656 GCs; 160 new targets come from FORS2 and 73 targets from
GMOS-S.
A typical velocity uncertainty is 50 km/s, while it can be considerably higher for the faintest targets, which have a brightness of about $m_R \approx 22.8$.

\section{Velocity dispersions}
\begin{figure}
\centering
\resizebox{\hsize}{!}{
\includegraphics[width=0.4\textwidth]{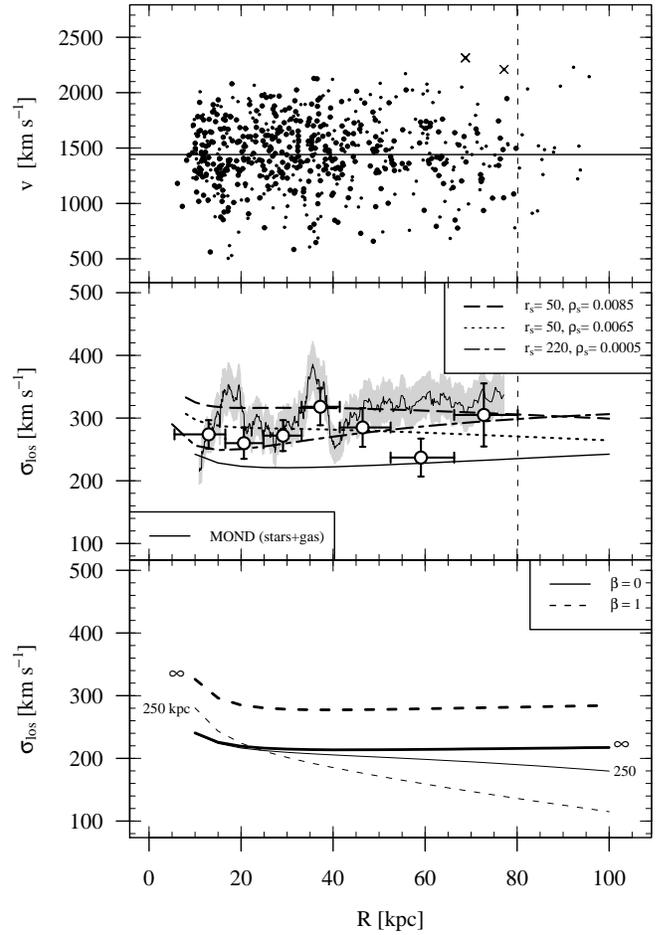}}
\caption{
Kinematics of GCs around NGC\,1399.
The upper panel shows the radial velocities of our total sample versus the radial distance,
where 80 kpc correspond to 14.5 arcmin.
The two crosses are outlier velocities that we have discarded, and the larger
points show objects from the ``safe'' sample.
The horizontal line marks the systemic velocity, and
the vertical dashed line indicates the radial limit
used for the ``safe'' sample.
The middle panel shows the velocity dispersion profile.
The thin solid line with shaded region shows the result derived for the full GC sample along
with its uncertainties, while the circles with error bars show the ``safe'' sample.
The curves represent various dynamical model fits (isotropic), as described in the text,
and with parameters summarized in the legend.
The lower solid line represents the expectation from MOND under isotropy.
The lower panel shows in addition fully radial MOND models (without gas) with a cut-off radius of
250 kpc.
}
\label{fig:dispersion}
\end{figure}

Converting measured velocities to  velocity dispersion is not straightforward.
Fluctuations may arise from inhomogeneous sampling or from small numbers of outliers
(stemming e.g. from erroneous velocities or contaminants belonging to other galaxies---in our case
 NGC\,1404, which is at a projected distance of 54 kpc). Therefore we define a ``safe'' subsample, where the objects have distances greater than 3 arcmin to NGC\,1404, velocity uncertainties smaller than
75 km/s, and magnitudes in the range $21 < m_R < 22.35$.
We calculate dispersions
by applying the maximum likelihood estimator of Pryor \& Meylan\,(\cite{pryor93}), with
the individual velocity uncertainties as weights.
For the full sample,
the dispersion was
calculated by moving bins, each bin containing 60 objects with a step of one object.
For the ``safe'' sample, we calculate the dispersion in independent bins.
The sample velocities and derived dispersions are shown in Fig.~\ref{fig:dispersion}.

The velocity dispersion is approximately constant out to
80\,kpc, thus continuing the trend found in Paper\,I.
The various curves refer to different models, which are explained in the
following sections.

\section{Models}
In Jeans models that only use velocity dispersions and not higher moments of the velocity distribution, the mass distribution and the orbital anisotropy of the tracer population are degenerate. 
Here we mainly consider isotropic models, which are good
approximations at least to the entirety of those clusters  that were found in Paper\,I to best
explain the GC dispersions
(see also
C{\^o}t{\'e} et al.~\cite{cote01}). In the MOND discussion,  we also give fully radial models (which are not realistic) as a guide to how a radial anisotropy affects the results.

We model the projected velocity dispersions under isotropy according to

\begin{equation}
\centering
\sigma_{los}^2(R)=\frac{2}{N(R)}\int_{R}^{\infty}\frac{\sqrt{r^2-R^2}}{r^2}\ell(r) M(r) dr\; ,
\end{equation}
where $N(R)$ is the projected number density of the tracer population,
$\sigma_{\rm{los}}$ the line--of--sight velocity dispersion to
be compared to our observed values,
$r$ is the radial distance from the center and $R$ the projected distance, $\ell$ the
spatial (i.e.,~three--dimensional) density of the GCs,
and $M(r)$ the enclosed mass. 
 See e.g.,~Mamon \&
{\L}okas (\cite{mamon05}) and van der Marel (\cite{marel93}) for more general analytical
solutions of the Jeans equation.

The surface number densities of GCs are adopted from Bassino et al.~(\cite{bassino06})
 as 
the sum of the number densities for metal-rich and metal-poor clusters, resulting in 
the 3D density
\begin{equation}
\ell(r)  = \frac{N_{0}}{R_0}
\frac{1}{\mathcal{B} \left(\frac{1}{2},\alpha \right)}
\cdot
\left( 1 +
\left(\frac{r}{R_{\mathrm{0}}}\right)^2 \right)^{-\left(\alpha+\frac{1}{2}\right)}\,
\label{eq:depropl}
\label{eq:corepl}
\end{equation}
with  the values
$N_0=35.54$,
$R_0=1\farcm74$,
$\alpha= 0.84$,
and the Beta-function $\mathcal{B} \left(\frac{1}{2},\alpha \right) = 2.22$.

As a model for the luminous matter, we adopt the power law given by
Dirsch et al.~(\cite{dirsch03}) resulting from the wide-field photometry of NGC\,1399
together with a  (constant) stellar mass to light ratio,
which we fix to ${M/L}_{\rm{R}}=5.5$, following Paper\,I.  This
is about the maximum value, which is found for early-type galaxies
(see also Saglia et al.~\cite{saglia00}, Bell et al.~\cite{bell03}).

For the shape of the dark halo, we assume an NFW type halo, which for the inner 40\,kpc already gave a good
description (Paper\,I).
 The cumulative mass is

\begin{equation}
\centering
  M_{\mathrm{dark}}= 4\pi\cdot\varrho_{\mathrm{s}}\cdot
  r_{\mathrm{s}}^3\cdot\Bigg(
  \ln\Big(1+\frac{r}{r_{\mathrm{s}}}\Big)
  -\frac{\frac{r}{r_{\mathrm{s}}}}
  {1+\frac{r}{r_{\mathrm{s}}}}
  \Bigg)\;,
\label{eq:nfw}
\end{equation}
where $\varrho_{\mathrm{s}}$ and $r_{\mathrm{s}}$ are the characteristic density and radius, respectively. 

\section{Mass profiles}
The full sample can be reproduced well by  adding an NFW halo  
with $\varrho_{{s}}=0.0085 M_\odot/\rm{pc}^3$ and $r_{s} = 50\,\rm{kpc}$,
 the ``safe'' sample with $\varrho_{{s},0}=0.0065 M_\odot/\rm{pc}^3$ and $r_{s} = 50\,\rm{kpc}$
 (long-dashed and
 dashed line in Fig.~\ref{fig:dispersion}) 
to the mass of NGC 1399.  As Table\,\ref{masses} shows, the gas mass (which is included
in the halo) is negligible within 80 kpc
with respect to the stellar mass and unimportant within the virial radius with respect to the dark
halo.

 The uncertainties are mostly of a systematic nature and
difficult to quantify, but these two halos probably embrace the acceptable values with a
preference near the ``safe'' sample  in the sense of the best-fit halos. 

\begin{figure}
\centering
\resizebox{\hsize}{!}
{\includegraphics[width=0.4\textwidth]{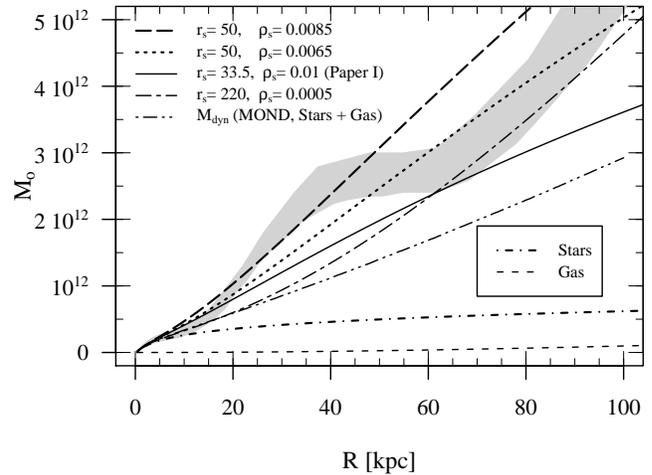}}
\caption{Comparison to X--ray measurements.
The grey area shows the range of the mass models presented by Paolillo
et al.~(\cite{paolillo02}). The short-dashed line refers to  our preferred halo, the long-dashed 
line to the halo derived from the full sample. The solid line is the halo
derived in Paper\,I. 
The stellar mass is shown as dashed-dotted line. The dynamical MOND mass is shown as the dot-dot-dashed line.}
\label{fig:fp}
\end{figure}

These halos can now be compared with the X-ray mass profile presented by Paolillo et al.~(\cite{paolillo02}).  Figure\,\ref{fig:fp} shows the  total
mass for the various models; i.e. the NFW halo includes dark matter, gas, and
the stellar matter outside NGC 1399; however, it is vastly dominated by 
dark matter.
 The overall agreement of the X-ray mass profile, particularly 
with the ``safe'' halo, is very good.
 There is no support however for the steep rise between 20 and 40\,kpc 
and the subsequent flat part, which would mean a strong decline in the mass density to almost zero. This feature has been already noted in the ASCA study of Ikebe et al.~(\cite{ikebe96}),
who interpreted it as the transition radius between the galaxy and the cluster potential.
Such a  mass profile would probably
be dynamically unstable, so we suspect that it is possibly a
consequence of azimuthally averaging over a disturbed intensity  field. 
The ``safe'' halo has a high concentration and  only marginally fits to the simulations of
Bullock et al.~(\cite{bullock01}), but better than the halo from Paper\,I. The virial radius is 670 kpc, the mass inside this radius is
$\mathrm 1.8 \times 10^{13} M_\odot$, and the concentration $\mathrm c = 13.4$.
 Until now, the only dynamical analysis of the Fornax cluster is that of Drinkwater et
al.~(\cite{drinkwater01}; hereafter DEA), who used galaxy velocities to constrain the cluster mass.\\
 To investigate what one would expect from the DEA mass for the projected velocities in our
radial range, we describe the DEA mass by an NFW profile (which includes dark matter and
baryons, but not NGC 1399) with the parameters $\varrho_{{s},0}=0.0005 M_\odot/\rm{pc}^3$ and $r_{s} = 220\,\rm{kpc}$, 
which reproduces  the solid line in Fig.~4 of DEA out to 1\,Mpc very well.
To compare this model to the GC dispersions in the inner regions of the halo,
we add in the stellar mass of NGC\,1399, and show the modeling result in Fig.~\ref{fig:dispersion}.
Although this halo model was not among our best fits, it is also not excluded, as it
seems to follow the velocity dispersions quite well.
However, it falls below the X-ray mass model at small radii (Fig.~\ref{fig:fp}).
The difference between the mass profile of DEA and our ``safe'' halo is listed in Table
\ref{masses}. The gas mass of Paolillo et al.~(\cite{paolillo02}) is also given. 
It is only known out to about 200 kpc, and thus is extrapolated to larger radii. 
Table~\ref{masses} shows that 
  the
agreement of our halos with 
the mass of DEA is  unsatisfactory. It becomes even worse at larger radii: 
at 1\,Mpc, the DEA mass is $6\pm 2\times 10^{13} M_\odot$ versus our
mass of $2.2 \times10^{13} M_\odot$ ($2.9 \times10^{13} M_\odot$ for the more massive halo).   
We conclude that, although our sample is already the largest one available, we are not able to
conclusively distinguish between different halos on a cluster-wide scale.  There is, however,
evidence from the data of Bergond et al.~(\cite{bergond07}) that our halo might be extrapolated
out to 200 kpc (Schuberth et al.~\cite{schuberth07}).

\begin{table}
\caption{The table lists the masses of different components or halos for two radii,
the larger one being the virial radius of the ``safe'' halo. All values are given in
units of $10^{12} M_\odot$. The stellar mass at 670 kpc is an extrapolation of the NGC1399
luminosity profile and serves as a proxy for the total stellar mass. The gas mass is taken from  Paolillo et al.~\cite{paolillo02} and is extrapolated beyond 200 kpc. The ``safe'' halo is
explained in the text. ``DEA-mass'' denotes the Fornax cluster mass according to Drinkwater
et al.~\cite{drinkwater01}.
}
\begin{center}
\begin{tabular}{ccccc}
\hline
Radius & stellar mass & gas mass & ``safe'' halo & DEA-mass \\
\hline
80 kpc & 0.6 & 0.04&  3.5 & 2.9 \\
670 kpc & 1.1 & 1.5 & 18 & 43\\
\hline

\end{tabular} \\
\end{center}
\label{masses}
\end{table}

\section{Modified Newtonian dynamics}
As an alternative to the CDM paradigm, Milgrom's MOND (e.g. Milgrom
 \cite{milgrom83}, Sanders \& McGaugh \cite{sanders02}) is mostly successful in accounting for the kinematics of
 disk galaxies and/or the baryonic Tully-Fisher relation (McGaugh \cite{mcgaugh05}). Little is known about
 early-type galaxies. Schuberth et al.~(\cite{schuberth06}) find that NGC\,4636 is indeed consistent with
 MONDian kinematics.
Also the declining velocity dispersion of NGC\,3379 (Romanowsky et al.~\cite{romanowsky03})
does not necessarily contradict MOND (Milgrom \& Sanders \cite{milgrom03}). However, on the scale of galaxy clusters, MOND apparently needs additional dark matter (e.g. Sanders \cite{sanders03}, Pointecouteau \& Silk \cite{pointecouteau05}).
It is therefore interesting to investigate whether NGC\,1399, as a central galaxy, is compatible with MOND
or not.

Assuming spherical symmetry, we apply the classic MOND recipe 
where $g$ and $g_N$ are the MOND acceleration and Newtonian acceleration, respectively,
$a_0$  a universal constant, and $\mu(g/a_0)$ is a 
 function interpolating between the Newtonian and the MONDian regimes. 

Following Famaey \& Binney (\cite{famaey05}), we adopt $ \mathrm \mu = x/(1+x)$ with $x = g/a_0$
(see also Zhao \& Famaey 
 \cite{zhao06}).
The circular
velocity curve then reads
\begin{equation}
\centering
\mathrm  V_{circ,M}^2  = \frac{V_{circ,N}^2}{2} + \sqrt{\frac{V_{circ,N}^4}{4} + V_{circ,N}^2 \cdot a_0 \cdot r\, } \; ,
\end{equation}
where $\mathrm V_{circ,N}$ is the Newtonian circular velocity. For $a_0$ we adopt the value
recommended by Famaey et al.~(\cite{famaey07}): $1.35 \times 10^{-8} \rm{cm\,s}^{-2}$.
 The analysis strongly depends on the shape
of the interpolation, since $g$ is comparable to $a_0$ even at 80 kpc. Adopting the Bekenstein 
(\cite{bekenstein04}) interpolation function would increase $V_{circ}$ significantly, but it 
does not fit spiral rotation curves.

 To calculate the MOND projected velocity dispersion, we used the standard
Jeans equation, with the Newtonian mass corresponding to the MONDian circular velocity.
The MOND modeling results are shown in Fig.~\ref{fig:dispersion}. 
Under isotropy, the MOND dispersions are significantly lower than the profile derived from the GC dynamics.

Fully radial models are shown in the lower panel. We caution that, in contrast to the isotropic case,  
the outer integration limit now gains importance. Integrating to infinity, the MOND dispersion could
reproduce the data. However, we do not have infinitely distant GCs. The extension of
the GCS is about 250 kpc (Bassino et al.~\cite{bassino06}). The model for cut-offs of 250 kpc shows that
a more realistic radial model  does not help in achieving better agreement with MOND.   
We therefore conclude that MOND does not explain the dynamics of
NGC\,1399 without additional dark matter.

Such an additional hypothetical dark halo must have a relatively small
core and  a rapidly declining density profile  in order to avoid an increasing circular velocity.
Without claiming precision, a halo like $\varrho_{add} = 0.05 M_\odot/\rm{pc}^3 (1+r/12)^{-4}$, where $r$ is in
kpc, would suffice. It corresponds to adding a dark mass of
$7 \times10^{11} M_\odot$ within a radius of 80\,kpc, comparable to the stellar mass.

The extension of this  MOND model to 1\,Mpc predicts a circular velocity corresponding to a
Newtonian mass of $4.5 \times 10^{13} M_\odot$ -- which agrees 
 better with the DEA mass.  We also caution that the MOND dispersion at a radius of about 200 kpc has not
yet been ruled out (Schuberth et al.~\cite{schuberth07}). Moreover, the amount of gas beyond 200 kpc is not well-constrained.
 The necessary dark halo  does not  falsify
 MOND, but probably reflects the general finding that MOND fails in galaxy
 clusters without invoking a dark matter component. Sanders (\cite{sanders07}) proposes neutrinos; but since even very cold neutrinos are not expected to clump on scales below a few hundred kpc, this would not be a viable hypothesis for NGC\,1399. Other possibilities are molecules (but then how avoid star formation?)  or warm-hot intergalactic gas (e.g. Takei et al.~\cite{takei07}). Given that only 10\% of the baryons at low redshift have been  identified 
(e.g. Bregman \cite{bregman07}), the amount of gas needed is not in imbalance with the missing baryons. It would, however, be surprising to find
such an amount of gas. Future observations with increased sensitivity  might well potentially 
falsify this scenario.

\section{Conclusions}
We have used our new sample of about 660 velocities of GCs to constrain the dark halo of
 NGC\,1399 within 80 kpc. The resulting halo, assumed to be of the NFW type, is more concentrated than the halos
 from cosmological simulations, but marginally in agreement.
A more massive alternative halo, which would account for the total mass of the Fornax cluster
as estimated by Drinkwater et al.~(\cite{drinkwater01}),
may also be consistent with our data.
We are thus not able to conclusively constrain  the cluster-wide halo from probes within 80 kpc.  
We also consider the MOND case and find that MOND is not able to reproduce the observed
 GC kinematics under isotropy without invoking an additional dark matter component,
 as already found for other galaxy clusters. A realistic radial anisotropy does not
 change this conclusion. Moreover, only the isotropic model reproduces the X-ray mass well. The hypothetical dark halo must have a small core, and thus
 neutrinos do not seem to be  viable candidates. It may be in gaseous form.  Under our assumptions
(particularly the interpolation between the MONDian and the Newtonian regime), 
  MOND makes a strong prediction, which perhaps will be falsifiable in the future.

\begin{acknowledgements}
We acknowledge helpful discussions with Benoit Famaey and thank an anonymous referee for
clarifying comments. This work has been supported by the
Chilean Center of Astrophysics Conicyt FONDAP Nr. 15010003 and by the Deutsche Forschungsgemeinschaft (DFG project HI-855/2). AJR was also supported by NSF Grant AST-0507729.
\end{acknowledgements}

\bibliographystyle{aa}
\end{document}